\documentclass[%
 aip,
 amsmath,amssymb,
 reprint,%
]{revtex4-1}

\usepackage{graphicx}
\usepackage{dcolumn}
\usepackage{bm}

\usepackage[utf8]{inputenc}
\usepackage[T1]{fontenc}
\usepackage{mathptmx}
\usepackage{etoolbox}

\usepackage{color}

\newcommand{\textgx}[1]{\textcolor{black}{#1}}
\newcommand{\textgb}[1]{\textcolor{black}{#1}}
\newcommand{\textgr}[1]{\textcolor{black}{#1}}

\newcommand{\textgg}[1]{\textcolor{black}{#1}}

\newcommand{\textgm}[1]{\textcolor{black}{#1}}

\newcommand{\textgmm}[1]{\textcolor{black}{#1}}

\def\<{\langle}
\def\>{\rangle}

\makeatletter
\def\@email#1#2{%
 \endgroup
 \patchcmd{\titleblock@produce}
  {\frontmatter@RRAPformat}
  {\frontmatter@RRAPformat{\produce@RRAP{*#1\href{mailto:#2}{#2}}}\frontmatter@RRAPformat}
  {}{}
}%
\makeatother
\begin{document}

\preprint{AIP/123-QED}

\title {Rheological signatures of a glass-glass transition in an aging colloidal clay}

\author{Roberta Angelini}
\altaffiliation [Electronic mail: ] {roberta.angelini@cnr.it}
\affiliation{Institute for Complex Systems (ISC-CNR), Sede Sapienza, Piazzale Aldo Moro 2, 00185 Roma, Italy}

\author{Domenico Larobina}
\affiliation{Institute of Polymers, Composites, and Biomaterials, National Research Council (IPCB-CNR), P.le E. Fermi 1, 80055 Portici, NA, Italy}

\author{Barbara Ruzicka}
\affiliation{Institute for Complex Systems (ISC-CNR), Sede Sapienza, Piazzale Aldo Moro 2, 00185 Roma, Italy}

\author{Francesco Greco}
\affiliation{Department of Chemical, Materials and Production Engineering, University of Naples Federico II, P.le Tecchio 80, Napoli 80125, Italy}

\author{Raffaele Pastore}
\altaffiliation [Electronic mail: ] {raffaele.pastore@unina.it}
\affiliation{Department of Chemical, Materials and Production Engineering, University of Naples Federico II, P.le Tecchio 80, Napoli 80125, Italy}

\date{\today}

\begin{abstract}
The occurrence of non-equilibrium transitions between arrested states has recently emerged
as an intriguing issue in the field of soft glassy materials.
The existence of one such transition has been suggested for aging colloidal clays (Laponite\textsuperscript{\textregistered} suspensions) at \textgx{weight concentration 3.0 \%},
although \textgg{further experimental evidences are necessary to validate this scenario}.
Here, we 
test the occurrence 
of this transition \textgm{for spontaneously aged  (non-rejuvenated) samples}, by exploiting the rheological tools of Dynamical Mechanical Analysis.
On imposing consecutive compression cycles to differently aged clay suspensions, we find 
that a quite abrupt change of rheological parameters occurs for ages around three days.
For the Young and elastic moduli, the change with the waiting time is essentially independent from the
deformation rate, whereas other "fluid-like" properties, such as the loss modulus, do clearly display some rate dependence.
We \textgg {also} show that the crossover identified by rheology coincides with deviations
of the relaxation time (obtained through X-Ray Photon Correlation Spectroscopy) 
from its expected monotonic increase with aging.
Thus, our results robustly support the existence of a glass-glass transition in aging colloidal clays,
highlighting characteristic features of their viscoelastic behaviour. 
 
\end{abstract}

\maketitle

\section{Introduction}
In recent years, soft materials have emerged as a privileged stage to unveil rich and unconventional "phase diagrams", with out-of-equilibrium and time-dependent features, including transitions between seemingly arrested states~\cite{AndersonNat2002, PhamScience2002, EckertPRL2002, RuzickaNatMat2011, AngeliniNC2014}. 
Colloidal clays have been largely exploited as good model-systems to address these topics~\cite{MichotPNAS2006,ShalkevichLang2007,
MouradJPCB2009,RuzickaSM2011}; these systems have also been proven useful as self-assembling materials~\cite{GlotzerNatMat2007}, thanks to their anisotropic shape and complex interactions.

Among those colloidal materials, an industrial synthetic clay, Laponite\textsuperscript{\textregistered}, is today considered a privileged system to spontaneously build up locally anisotropic structures~\cite{RuzickaSM2011, RuzickaNatMat2011}.  Laponite\textsuperscript{\textregistered} in water is a charged colloidal suspensions of disc-shaped particles of nanometric dimensions, with an aspect ratio 25:1, having a net negative charge on the faces and a positive one on the rims. The anisotropic particle shape together with the presence of competitive repulsive and attractive Coulombic interactions and with their directionality are at the origin of very intriguing aging phenomena, driving the suspension from a liquid state towards different arrested states,  depending on clay and salt concentration~\cite{RuzickaSM2011}. Since the
pioneering work of Mourchid et al.~\cite{MourchidLang1998}, the state diagram of such suspensions has been very debated and the arrested states have been alternatively distinguished in gels and
glasses~\cite{MongondryJCIS2005,TanakaPRE2004,
JabbariPRE2008,RuzickaSM2011}. 
The structure and relaxation dynamics of Laponite\textsuperscript{\textregistered} during aging have been deeply investigated through different experimental techniques such as light~\cite{NicolaiJCIS2001, BellourPRE2003, RuzickaPRL2004, SchosselerPRE2006, RuzickaLang2006, JabbariPRL2007, PujalaSM2012, SahaSM2014}, X-ray~\cite{BandyopadhyayPRL2004, RuzickaPRE2008, RuzickaPRL2010, RuzickaNatMat2011, HansenSM2013, AngeliniSM2013, AngeliniNC2014, AngeliniCSA2014, AngeliniCSA2015} and neutron scattering~\cite{BhatiaLang2003, TudiscaRSC2012,TudiscaPRE2014, MarquesSM2015}, rheology~\cite{JoshiRSPA2008, AngeliniNC2014, SumanLangmuir2018}, 
 as well as through theory and simulations~\cite{DijkstraPRL1995, KutterJCP2000, TrizacJPCM2002, OdriozolaPRE2004, MossaJCP2007, JonssonLang2008, DelhormeSM2012, JabbariSciRep2013} and display a range of intriguing features, including signatures of "anomalous aging", as observed in a variety of soft glassy materials~\cite{cipelletti2000universal,duri2006length,zia2014micro,pastore2021elastic,pastore2020anomalous,buzzaccaro2015spatially}.

Focusing on salt-free water conditions, combination of experiments, theory and simulations suggests that \textgx{gel and glass} are respectively obtained  below and above a clay concentration of C$_w=$2.0 \%~\cite{RuzickaSM2011}. 
\textgx{Below} C$_w$ = 2.0 \%, \textgx{where} attractive interactions are thought to dominate, an "empty liquid" and an equilibrium gel have been located for C$_w$ $<$ 1.0 \% and 1.0 \% $<$ C$_w$ $<$ 2.0 \%, respectively~\cite{RuzickaNatMat2011}.
On the other side, \textgx{for C$_w \geq$ 2.0 \%, the formation of a Wigner glass (WG)~\cite{BonnEL1999, RuzickaPRL2010},  i.e. an arrested state formed by disconnected particles stabilized by electrostatic repulsions, has been proposed}.
\textgm{At those relatively high concentrations, contrasting opinions about the nature of the arrested states (whether glass or gel) do exist~\cite{BonnEL1999,TanakaPRE2004,JabbariPRE2008, RuzickaPRE2008,JoshiRSPA2008,RuzickaPRL2010,RuzickaNatMat2011,RuzickaSM2011,AngeliniNC2014,MarquesJPC2017,SumanLangmuir2018}. Here, drawing on the “glassy” scenario~\cite{TanakaPRE2004,JoshiRSPA2008,RuzickaNatMat2011,AngeliniNC2014}, we emphasize that,} even in the WG, attractive interactions may play a relevant role, possibly leading to a slow evolution of local particle configurations with waiting time.
\textgm{This picture is based on experiments at high concentration, C$_w$ = 3.0 \%, which have been interpreted in terms of a spontaneous glass-glass transition: with increasing waiting time, the system spontaneously evolves from the WG to a second glassy state~\cite{AngeliniNC2014}.
It is worth noticing that glass-glass transitions~\cite{PhamScience2002, EckertPRL2002, ChenSci2003,MayerNatMat2008} are quite rare to be found, especially at ambient conditions, and are usually driven by a change of some external parameter.}

\textgm{The possibility of a spontaneous glass-glass transition in Laponite\textsuperscript{\textregistered} suspensions is mainly supported by the following experimental results~\cite{AngeliniNC2014}:
1) for glassy samples rejuvenated by pumping them in a syringe, the late decay of the dynamic correlation function shows a crossover from a stretched to a compressed exponential,
occurring around three days after rejuvenation;
2) in a similar range of time after rejuvenation, the position of the peak of the structure factor $S(Q)$  shifts of about 10\%;
3) for spontaneously aged (non-rejuvenated) samples, a counterpart of the above mentioned results consists in a jump of the storage G$^{'}(\omega)$ and loss G$^{''}(\omega)$ moduli, measured under oscillatory shear, for waiting time around three days after sample preparation.}

\textgm{Other important insights of~\cite{AngeliniNC2014} arise from simulations of a stylized numerical model. The idea is that the equilibrium structure obtained through Monte Carlo (MC) dynamics at different temperatures may give some hints on the out-of-equilibrium structure of real Laponite\textsuperscript{\textregistered} suspensions at different waiting times. The simulations 
show that, around a given MC temperature, the local structure of platelets tends to preferentially rearrange into disconnected T-shaped geometries, with the entire system stabilized by orientational attractions: accordingly, the second glassy state at long waiting times has been called Disconnected House of Cards (DHOC)~\cite{AngeliniNC2014}.
Such a transition from a WG to a DHOC is clearly detected by visualizing the local particle configuration in direct space, which is, of course, straightforward in simulations, but experimentally unfeasible with state-of-the-art techniques for actual Laponite\textsuperscript{\textregistered} suspensions.
Thus, the structural changes observed in simulations cannot be directly validated in experiments.
However, simulations also show that the direct-space transition from WG to DHOC is accompanied, in reciprocal (Fourier) space, by a change of the structure factor $S(Q)$, which is qualitatively compatible with that observed in experiments (see point 2). This result strongly supports the possibility that the structural changes observed in simulations mirror those occurring in experiments. However, all of this is still far from being an ultimate proof, as no univocal relation exists, in general, between $S(Q)$ and direct space configurations (essentially, the same $S(Q)$ may correspond to different types of direct space configurations).}

\textgm{Thus, while~\cite{AngeliniNC2014} has the great merit to firstly highlight the presence of the transition, it should be clear that 
much is left to know about its nature, 
and that, in the impossibility of experimentally monitoring the local structure in direct space, the only feasible way to probe the proposed transition consists in collecting further "indirect" evidences, in addition to the three ones provided in~\cite{AngeliniNC2014}}. 

This is exactly the goal of this work, where the viscoelastic response upon compression of Laponite\textsuperscript{\textregistered} suspensions at C$_w$ = 3.0 \% in salt-free conditions  is investigated in depth through Dynamical Mechanical Analysis for a range of waiting times, and compared with X-Ray Photon Correlation Spectroscopy (XPCS) measurements \textgm{of autocorrelation functions}, to follow the evolution of the system dynamics across the proposed glass-glass transition. 

\textgm{While the shear rheology of Laponite\textsuperscript{\textregistered} has been investigated in several papers (for example~\cite{mourchid1995phase,abou2003nonlinear,JoshiRSPA2008,labanda2008effect,AngeliniNC2014,lapasin2017rheology,SumanLangmuir2018}) under a variety of conditions, DMA tests have never been used, to the best of our knowledge.
Notably, our rheological characterizations are performed both in a non-linear and linear regime, trough stress-strain compression cycles and frequency sweep tests, respecively, and cover waiting times significantly larger than in~\cite{AngeliniNC2014}.}

\textgm{It is worth remarking that here we chose on purpose to focus on non-rejuvenated samples, for which only one evidence of the transition is provided in~\cite{AngeliniNC2014} (see point 3 above). When dealing with rejuvenated samples, in fact, the outcomes may strongly depend on the type and the details of the adopted rejuvenation protocol. Furthermore, rejuvenation protocols, such as pumping the suspensions in a syringe like in~\cite{AngeliniNC2014}, hardly allow for fine control of the relevant parameters (e.g. the shear rate). Thus, we firmly believe that non-rejuvenated samples are the most appropriate standard to probe the proposed transition.}

\textgm{
Here we do provide the following new experimental evidences supporting the existence of a transition for ages just around 2-3 days:
a)	a jump in the elastic modulus and in the residual strain measured through DMA upon compression;
b)	a jump in the storage and loss moduli measured upon compression, through frequency sweep oscillatory rheology;
c)	an anomalous trend of the relaxation time and of the strecthing exponent measured through XPCS.}


\begin{figure}[t]
\centering
\includegraphics[width=1.05\linewidth]{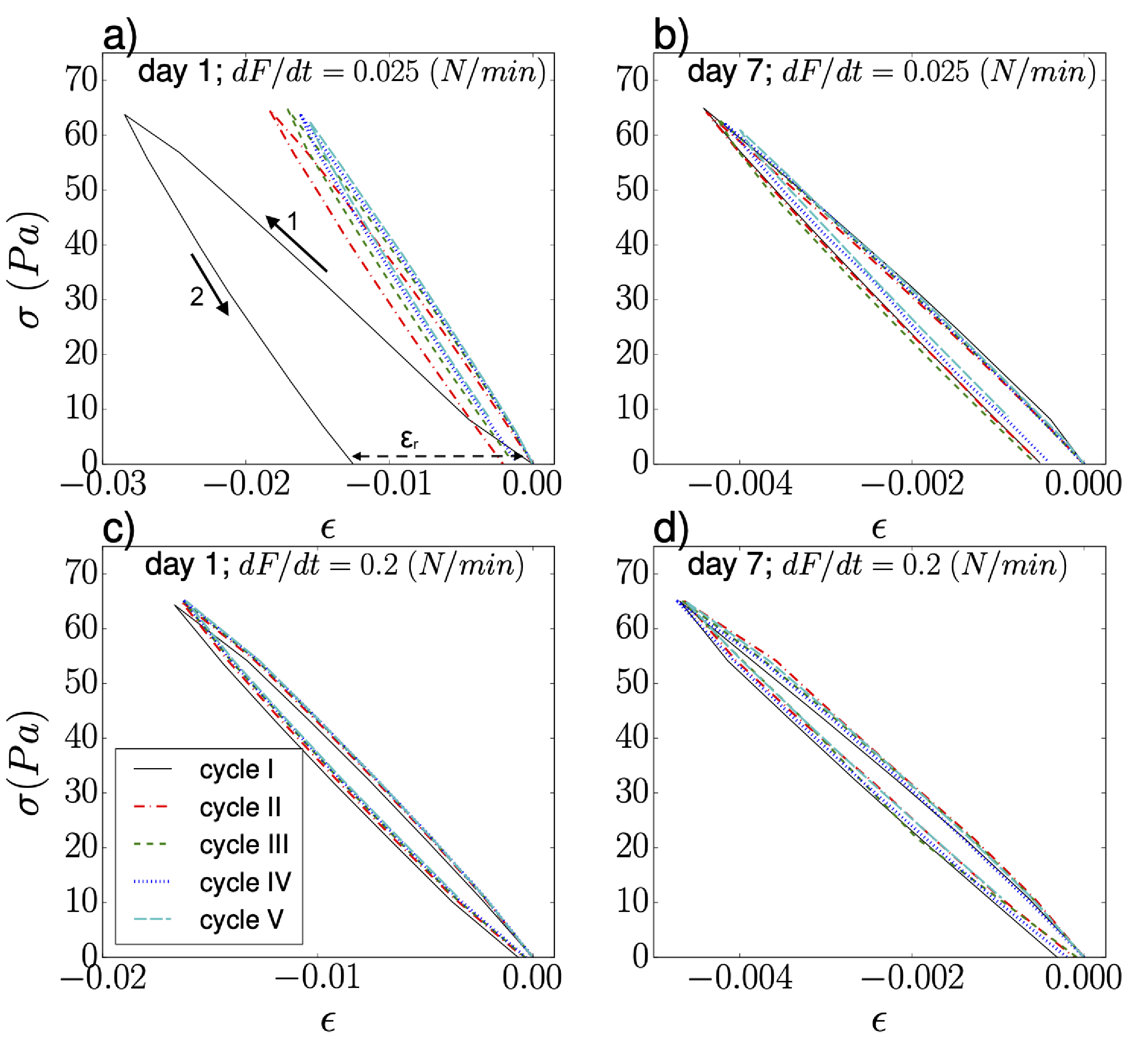}
\caption{Stress-strain curves for five consecutive DMA cycles for samples  
aged for 1 day and 7 days, submitted to force rates dF/dt = 0.025 N/min (a,b)  and 0.2 N/min (c,d).  
For each cycle, a backward shift has been applied, equal 
to the residual strain at the end of the previous cycle.
\textgm{For the exemplificative case of the first cycle in panel (a), the arrows indicate the compression (1) and decompression (2) ramps of the cycle, and the residual strain $\epsilon_r$.} 
}
\label{fig1}
\end{figure}

\begin{figure}[t]
\centering
\includegraphics[width=0.8\linewidth]{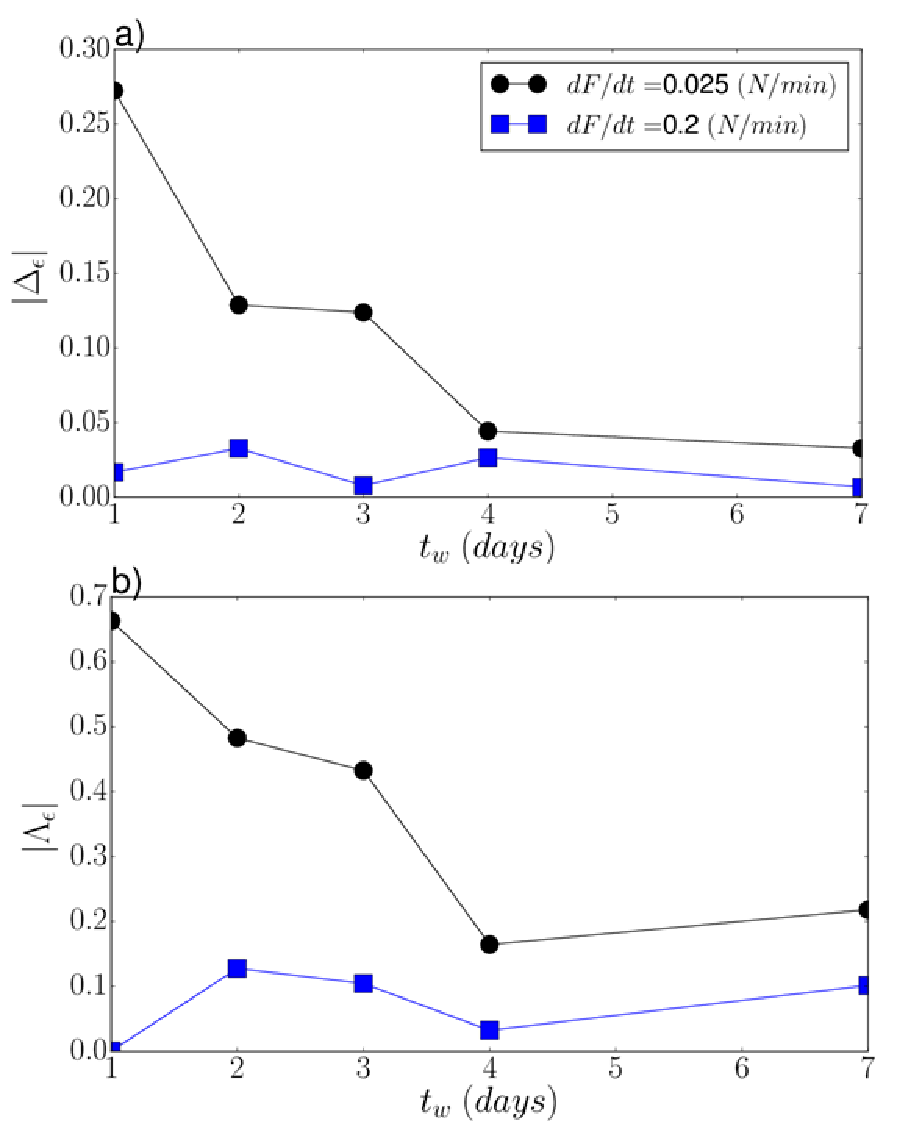}
\caption{(a) $\Delta_{\epsilon}$ and (b) $\Lambda_{\epsilon}$ as a function of the waiting time \textgb{for the smallest and the largest force rates}.
}
\label{fig2}
\end{figure}

\section{Experimental method}

\subsection{Sample preparation}
Laponite\textsuperscript{\textregistered} is a synthetic smectite clay, belonging to the 2:1 phillosilicate family, with its unitary cell consisting of one octahedral layer of six magnesium ions, sandwiched between two layers of four tetrahedrally coordinated silicon atom groups, in the presence of four oxydril groups (OH) and interlayer sodium cations. The substitution of some divalent magnesium ions in the octahedral sheet by monovalent lithium ions,  provides  a negative charge of -0.7$e$ to the unit cell and a molecular formula $Na^{+0.7}[(Si_{8}Mg_{5.5}Li_{0.3})O_{20}(OH)_{4}]^{-0.7}$~\cite{LalJApplCryst2000}; moreover in aqueous solution the dissociation of $OH^-$ ions from the rims  leads to their positive charge and raises the pH of the solution~\cite{TawariJCIS2001}. The unitary cell repeated about 1500 times in two dimensions, forms a Laponite disc with a diameter of 25 nm and 1 nm thick~\cite{RuzickaSM2011} with a uniformly distributed negative charge on the surface, around several hundred unit charge $e$ and positive on the rims, ten times lower for a salt-free system, as in our case~\cite{JabbariPRE2008, MartinPRE2002, TawariJCIS2001}.

The sample was prepared at a weight concentrations C$_w$ = 3.0 \%,  as extensively described in~\cite{RuzickaSM2011}: the oven-dried Laponite\textsuperscript{\textregistered} RD manufactured by Laporte Ltd was dispersed in ultra pure deionized water ($C_s\approx 1 \times 10^{-4}$ M) and kept under stirring for 30 minutes. Soon after, it was filtered through a 0.45 $\mu$m pore size Millipore filter in a Petri dish closed and sealed and the time at which the suspension was filtered was considered the origin of the waiting time ($t_w$ = 0). \textgx{From this moment on, the sample starts aging up to undergoing a liquid-glass transition at a time of the order of few hours~\cite{MarquesJPC2017}}. Cylindrical specimens of about 15 mm diameter and 5 mm thickness were obtained through an handmade shaper, \textgm{as shown in Fig. S1a of the supplementary material.}
It is worth remarking \textgm{once again here} that, at variance with~\cite{AngeliniNC2014}, no rejuvenation protocol is applied in any of our experiments.

\subsection{Dynamical mechanical measurements}
The viscoelastic response of Laponite\textsuperscript{\textregistered} suspensions has been investigated through a Dynamic Mechanical Analyzer (DMA) Model Q800 from TA Instruments. 
The cylindrical samples of aqueous susensions of Laponite\textsuperscript{\textregistered} at C$_w$ = 3.0 \% were placed on the plate, kept at 25 $^o$C and measured in compression configuration with controlled force\textgm{, as shown in Fig. S1b and c of the supplementary material.}

These kind of mechanical tests allow to measure the elastic modulus under compression, E, in analogy with measurements performed in shear mode, \textgx{which provide} the shear modulus (G). Two types of tests were performed: stress-strain and frequency sweeps.

In the {\it stress-strain} tests, a ramp force has been performed up to a maximum compression value F$_{max}$ = 0.01 N, followed by
a corresponding decreasing ramp down to the vanishing of the  force; 
five consecutive cycles of such ramp-up/ramp-down protocol have been performed for each experiment. 
\textgb{Due to an instrumental issue, the last cycle of each experiments is not fully complete (i.e. zero force condition is not completely restored at the end of each experiments).}

The stress, $\sigma$, given by $\sigma ={\frac {F_n}{A}}$,  
where $F_n$ is the compression force and $A$ is the \textgb{nominal} surface of the sample, 
and the corresponding strain \textgb{$\epsilon=\frac{L-L_0}{L_0}$ have been measured, with
$L$ and $L_0$ being the actual and the initial height of the sample.}

Different force rate, $dF_n/dt$, have been applied in the range (0.025 $\div$ 0.2) N/min, and measurements were performed at different waiting times, $t_w$, from one day up to one week.

\textgm{For each investigated waiting time and force rate, a new sample was used.
Evaporation effects over the duration of a measurement (4 minutes at most) were ruled out through water evaporation tests as a function of time, as reported in Fig. S2.}
\textgm{For any investigated waiting time and force rate, three independent measurements were performed. See Fig.S3 for some examples of measurements from replicated experiments.}

In the  {\it frequency sweep} tests an oscillatory experiment has been carried out with a sinusoidal compression \textgb{stress $\sigma$
\begin{equation}
\sigma(t)=\sigma_{0}sin(\omega t)
\end{equation}
and a strain $\epsilon$ that, for a viscoelastic material, presents a phase lag $\delta$
\begin{equation}
\epsilon(t)=\sigma_{0}J'sin(\omega t)+ \sigma_{0}J''cos(\omega t)
\end{equation}
with $J'= \epsilon _0/\sigma_0\cos \delta$, $J''=\epsilon _0/\sigma_0\sin \delta$,
where $\sigma_{0}$ is a fixed compression stress, $\omega$ = 2$\pi f$ with $f$ the frequency, and J$'$ and J$''$ are the compliances. $\sigma_0=28.3$ Pa has been appropriately selected in the 
linear regime, drawing on preliminary amplitude-sweep tests at $1Hz$ at each investigated age. 
In the linear regime, the moduli are connected by the following equations: 
$E'=\frac{J'}{J'^2+J''^2}$ and $E''=\frac{J''}{J'^2+J''^2}$~\cite{ferry1980viscoelastic}}, where $E'$ is the storage modulus of the material in-phase with the strain and $E''$ the loss modulus out of phase with it.
Similarly to the shear storage, G$'$, and loss, G$''$, moduli, E$'$ and E$''$ are related to the
material`s ability to store energy elastically and dissipate stress through heat, respectively. 
\textgm{Frequency sweep tests were performed at the same waiting times as for stress-strain cycle tests, and a new sample was used for each investigated waiting time. 
Measurements at a few values of the waiting time were repeated on further samples to check reproducibility.}

\begin{figure}[t!!]
\centering
\includegraphics[width=1.05\linewidth]{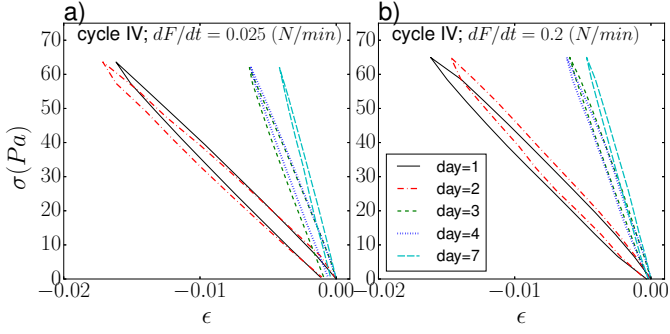}
\caption{Stress-strain curves for the fourth DMA cycle for samples submitted to force rates (a) dF/dt = 0.025 N/min and  (b) dF/dt = 0.2 N/min 
and aged for different waiting times, as indicated.
}
\label{fig3}
\end{figure}
%
%
\begin{figure}[t!!]
\centering
\includegraphics[width=1.0\linewidth]{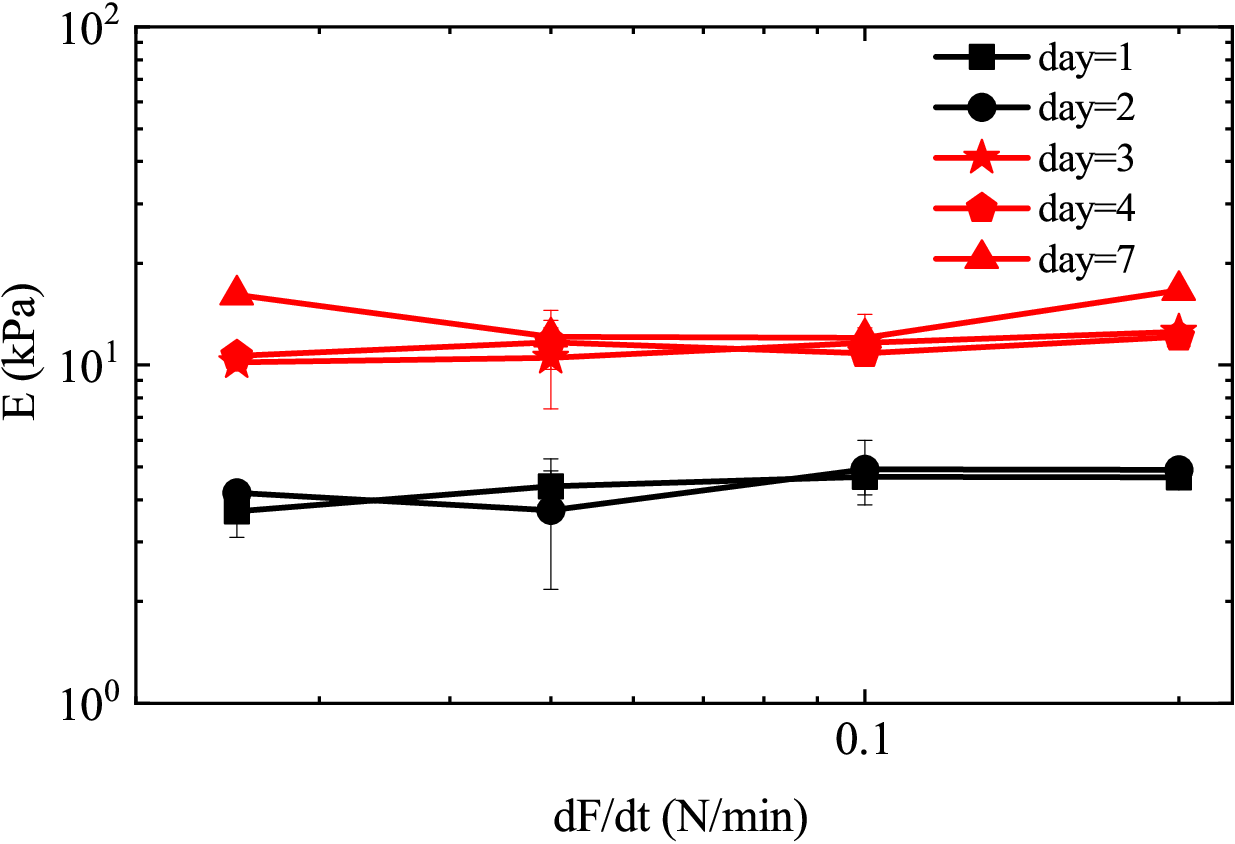}
\caption{
Young modulus $E$, estimated from a linear fit to the initial part of datasets like those in Fig.~\ref{fig2} averaged over five DMA cycles \textgm{and over independently measured samples},
as a function of the normal force rate at different waiting times.}
\label{fig4}
\end{figure}

\subsection{XPCS measurements}
\label{sec:XPCSmethods}
\textgx{XPCS measurements were performed at the ID10 beamline at
the European Synchrotron Radiation Facility (ESRF, Grenoble,
France). Using an incident
partially coherent X-ray beam with energy fixed at 8 keV a series of
scattering images was recorded by using a charged coupled device (CCD) and averaged intensity autocorrelation
function 
\begin{equation}
g_2(Q, t)=\frac{\langle\langle
I(Q,t_0)I(Q,t_0+t)\rangle_p \rangle}{\langle \langle
I(Q,t_0)\rangle_p \rangle^2}
\end{equation}
was calculated by using a standard multiple $\tau$
algorithm~\cite{MadsenNJP2010}}.
Here, $\langle...\rangle_p$ is the
ensemble average over the detector pixels mapping onto a single Q
value and $\langle...\rangle$ is the temporal average over $t_0$,


\textgm{Our XPCS results include a dateset from a previous work~\cite{AngeliniSM2013} and two new datasets targeted to a restricted and finely resolved range of waiting times, just within the time-window of the expected glass-glass transition.
The three datasets correspond to three different samples that have been prepared at the same time and tested for different time-spans within the same overall beamtime, by repeatedly switching the samples within the apparatus.
Such a protocol has been specifically designed to check reproducibility while probing the largest waiting-time range (up to 3.11 days) within the limited available beamtime. 
Indeed, within the given beamtime, it would not have been possible to cover the same waiting-time range with "in-series" measurements for three distinct samples.
To further support the reproducibility of our results, we also analysed a new dataset on Laponite\textsuperscript{\textregistered} suspension, always at C$_w$=3.0\%, but in deuterated water D$_2$O, from an experiment originally designed to make a comparison with neutron spin echo measurements~\cite{MarquesSM2015}.}

%
%
%
\textgr{\section{Results and discussion}}
\textgr{\subsection{Stress-strain cycles}}

We start by giving an overview  \textgx{of} the stress-strain cyclic tests.
\textgb{In the four panels of Fig.~\ref{fig1}, we show $\sigma$-$\epsilon$ plots at the smallest \textgx{(1 day) and the largest (7 days) waiting times (columns)
and for the minimum (dF/dt  = 0.025 N/min) and maximum (dF/dt  = 0.2 N/min)} imposed force rates (rows).}
In each panel, all the five consecutive cycles have been included, with a proper shift of the corresponding curves \textgx{backward to zero strain at any starting of a cycle.} Shifting is indeed necessary to get a more neat representation of the experiments, as cycles do not "close", i.e. some detectable residual strain is often present at the end of each cycle. 
In all cases, an hysteresis is found in the material response, whose magnitude is almost unaffected by the cycle number, especially at the largest force rate. A single yet fully reproducible exception is found for the first cycle of the youngest sample at the lowest force rate (see panel a, for which the hysteretic behavior is much more marked and, eventually, the residual strain is much larger than for the following four cycles. 
Data show that, in the initial portion of any cycle, a linear stress-strain relation holds, thus allowing for the measurement of an effective elastic modulus, to be discussed below.
Finally, we notice that the maxima of the strain, attained at the end of the rising ramps, are significantly larger (by more than a factor 3 and independently from the force rate) for the young sample.
\textgm{Such an "age-stiffening" effect is, in fact, observed in many soft amorphous solids, including (physical) polymer gels and colloidal glasses (for example~\cite{zia2014micro,siviello2015analysis,johnson2019influence,pastore2020anomalous,panja2022controlling}), and is commonly ascribed to a compactification of the underlying structure.}

In order to better show to what extent consecutive cycles differ from each other, both in terms of maximum of the strain and of width of the hysteresis, we define: 
\begin{equation}
\label{eq:Delta}
\Delta_{\epsilon}=\frac{\epsilon_f^*-\epsilon_i^*} {\epsilon_f^*+\epsilon_i^*}
\end{equation}
where $\epsilon_f^*$ and $\epsilon_i^*$ are the maxima of the strains attained during  the final and initial cycles, respectively, and

\begin{equation}
\label{eq:Lambda}
\Lambda_{\epsilon}=\frac{\Gamma_f-\Gamma_i} {\Gamma_f+\Gamma_i}
\end{equation}
where $\Gamma_f$ and $\Gamma_i$ are the full-width at half-maximum of the final and the initial cycles, respectively.
\\
Figure~\ref{fig2} shows the absolute values of $\Delta_{\epsilon}$ and $\Lambda_{\epsilon}$ in panels a and b, respectively, as a function of waiting time and \textgb{for the smallest and the largest  force rates.
A clear trend is observed only at the smallest force rate, where both $|\Delta_{\epsilon}|$ and $|\Lambda_{\epsilon}|$ are found to decrease over the first three days.
By contrast, on increasing the force rates, the curves becomes essentially flat over the whole waiting time range, as exemplified  through the data at the largest $dF/dt$ in the figure.}

Overall, Fig.s~\ref{fig1} and~\ref{fig2} suggest that the material response is closely elastic and memoryless for all experiments, but for young samples at the lowest force rate. In the latter case, the response seems to be affected by the short-term history of the sample, with a more marked viscoelastic and, at the same time, weaker solid-like response, especially during the first DMA cycle. 
Such a behaviour is due to the material relaxation time
being comparable (in the young samples) with the characteristic time of the imposed deformation. In agreement with our observations, those times are expected to become \textgx{significantly different} (and, in turn, viscoelastic effects less relevant) both on increasing  waiting time and  force rate. Indeed, the relaxation time usually increases with the waiting time \textgx{~\cite{AngeliniSM2013}}. 
The present data suggest that, in the young samples, the interplay between structural relaxation and deformation rate results in an hardening induced by successive cycles.


To illustrate more neatly how the material response depends on waiting time and force rate, we now focus on the fourth DMA cycles (i.e. the last fully complete cycle of each experiment) , where any memory effect has become much less relevant (even for young samples at the smallest force rate). 
The curves obtained at all investigated waiting times are reported in Fig.~\ref{fig3}, at the smallest and the largest force rate in panels a and b, respectively.
It is quite evident that no qualitative difference appears by comparing the two panels.
In both cases, the initial slope of the curves tends to increase on aging. Such aging-induced-hardening shows, however, a quite abrupt "jump" with waiting time, as indicated by the clear clustering of the curves for $t_w$ $\leq$ 2 days and for $t_w$ $\geq$ 3 days, respectively, which is a first signature of a sudden crossover in the rheological properties taking place in between those time intervals.

To further characterize this distinctive behaviour, 
we show in Fig.~\ref{fig4} the effective Young modulus E as a function of the force rate at five different waiting times. 
At any waiting time and force rate, E was obtained by averaging the slope of the initial parts of each ramp-up over cycles\footnote{No relevant differences are observed between the so-averaged Young modulus and the one corresponding to the fourth cycles}, \textgm{and over independently measured samples (see Fig. S3a for some example of replicated measurements)}.

A very similar behaviour is observed for samples one-day and two-days old, with values of the Young modulus of the order of a few Pa. On the contrary, for waiting times higher than three days, a jump of E to values around 10 Pa is found. The existence of two different rheological behaviours is consistent with the occurrence of a glass-glass transition in Laponite$^{\textregistered}$ suspensions at C$_w$ = 3.0 \%. 

%
%
\begin{figure}[t]

\centering
\includegraphics[width=1.0\linewidth]{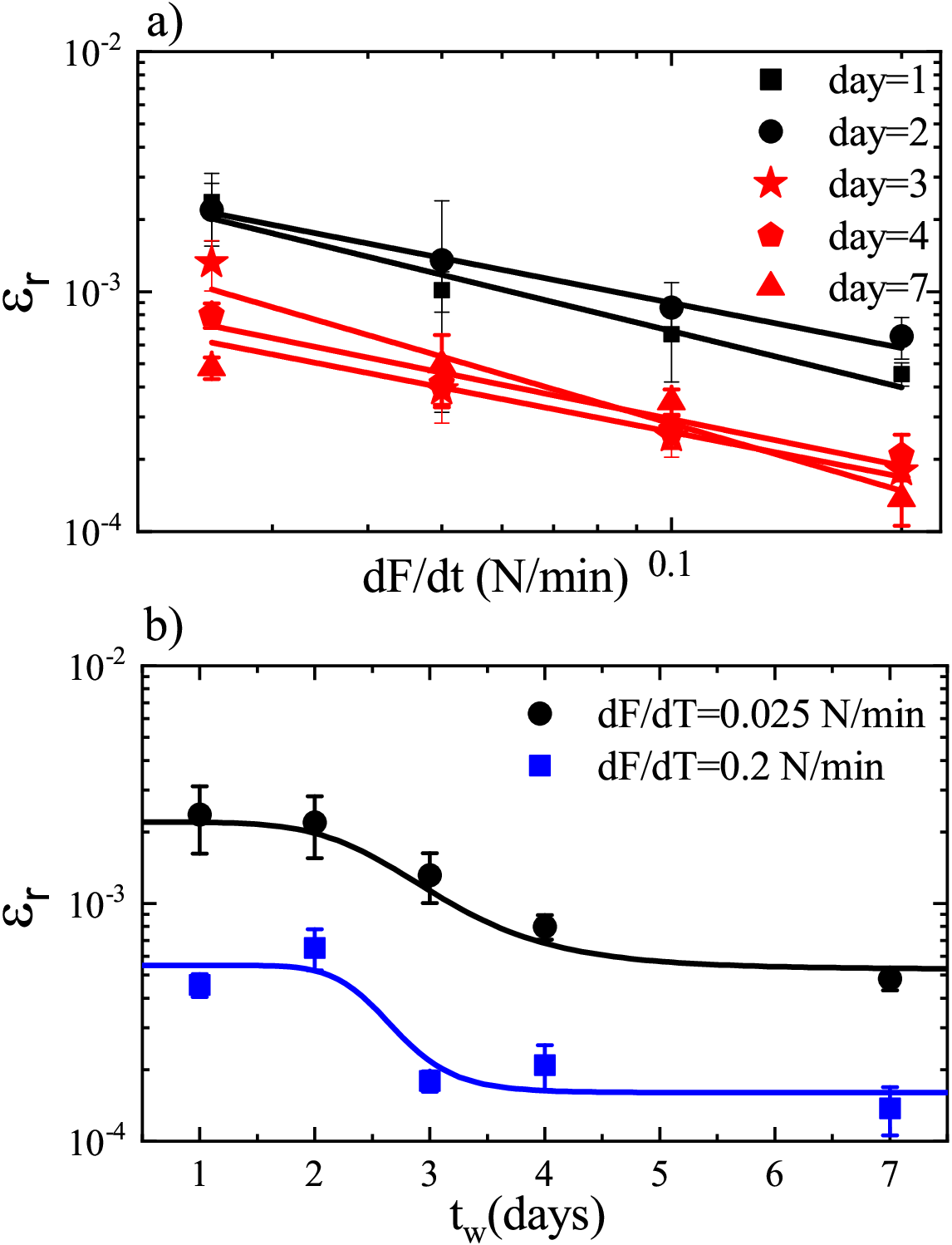}
\caption{Residual strain (a) as a function of the force rate at different waiting times, and (b) as a function of the waiting time for the smallest and the largest  force rates. Data are averaged over four DMA cycles \textgm{and over independently measured samples}. \textgm{Solid lines in both panels are guide to the eyes and should not be considered as suggestive of underlying functional forms}.
}
\label{fig5}
\end{figure}

To corroborate these findings, the residual strain $\epsilon_r$ \textgm{(determined as the strain when the stress goes back to zero at the end of each shifted cycle)}, averaged over four cycles \textgm{and over independently measured samples (see Fig. S3b) for some example of replicated measurements)}, is shown in Fig.~\ref{fig5}a as a function of force rate at different waiting times. Data are characterized by a linear decreasing trend and, again, by the existence of two different families of curves for samples measured before and after three days, reinforcing the idea that, in between two and three days, the system undergoes a transition towards a different glassy state. In Fig.~\ref{fig5}b), complementing Fig.~\ref{fig5}a,  the residual strain $\epsilon_r$ is shown as a function of waiting time, for \textgb{the smallest and the largest  force rates}. Curves display a tendency to plateau at low and high waiting times, with a clear deflection just \textgm{in between 2 and 3 days}.

\begin{figure}[t]
\centering
\includegraphics[width=1\linewidth]{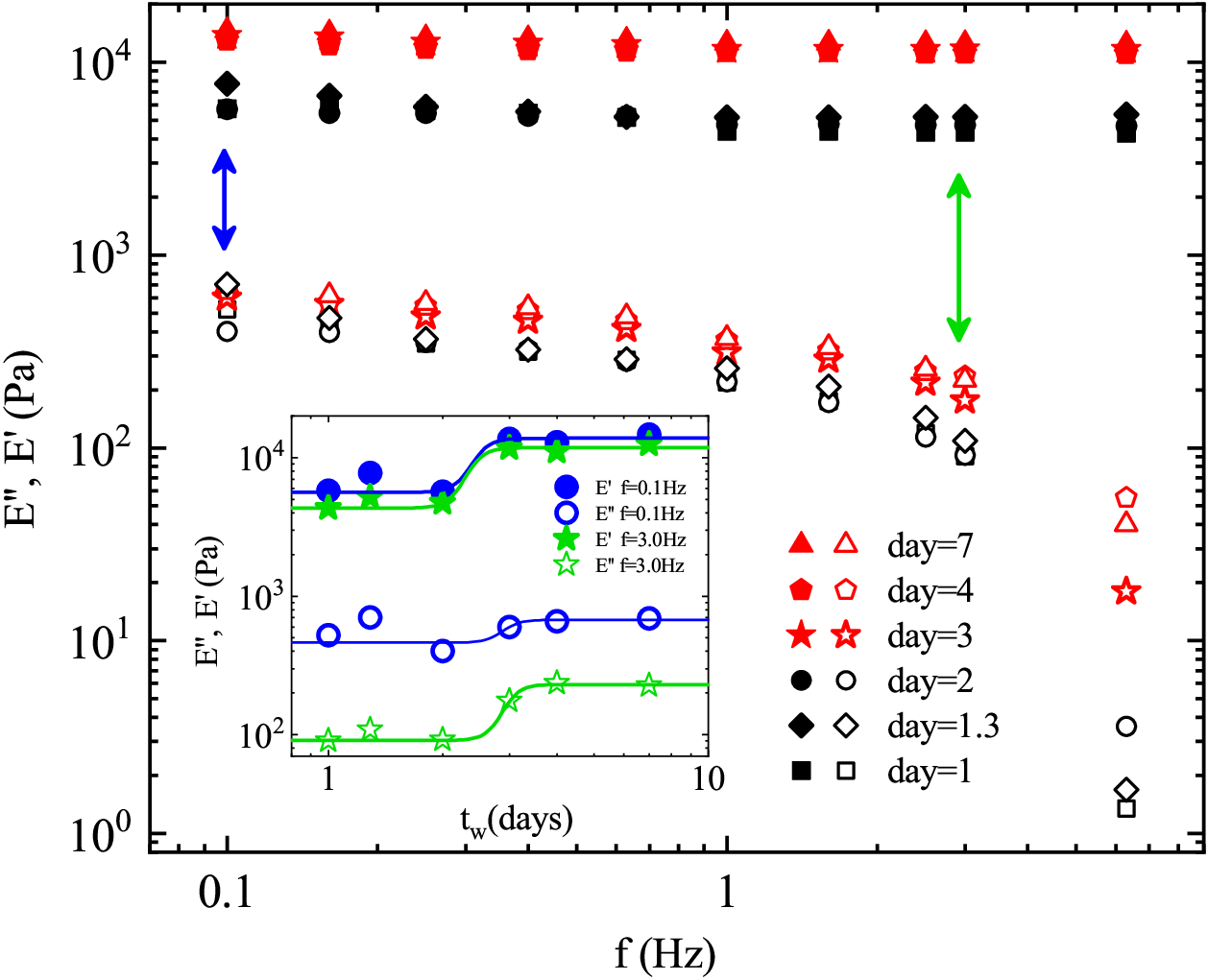}
\caption{Main: frequency dependence of the elastic and loss moduli, $E'(\omega)$ (full symbols) and $E''(\omega)$ (open symbols)
at different waiting times,  as measured in frequency sweep experiments. 
Inset: Elastic (a) and loss (b) moduli (from the frequency sweep experiment)
 as a function of the waiting time at the two frequency values  f = 0.1 Hz and f = 3.0 Hz indicated by the arrows. \textgm{Solid lines are guide to the eyes and should not be considered as suggestive of underlying functional forms.}
}
\label{fig6}
\end{figure}

\textgr{
\subsection{Frequency sweep tests}
}
To further explore the indicated transition, we now describe the results of oscillatory measurements at different waiting times, as shown in Fig.~\ref{fig6}. In all cases, the elastic modulus E$^{'}(\omega)$ is almost one order of magnitude larger than the loss modulus E$^{''}(\omega)$, confirming the solid-like nature of the system. Moreover, the elastic modulus E$^{'}(\omega)$ displays a frequency-independent behaviour over the probed frequency range, while E$^{''}(\omega)$ exhibits a decrease at high frequencies, indicating a change of the "liquid-like" properties of the system.  Even in this experiment, a jump of the moduli, more evident for the elastic modulus, arises. In fact, for ages up to roughly 2 days,
E$^{'}(\omega)$ is about $5 \cdot 10^3$ Pa, while  it is larger by a factor two/three above 3 days.
In the inset of Fig.~\ref{fig6}, the moduli are reported as a function of waiting time at two selected frequency 0.1 Hz and 3 Hz, as indicated by the arrows in  the main panel. The figure highlights that, for $E^{''}$, the jump is only evident at high frequency.


\begin{figure}[t]
\centering
\includegraphics[width=1\linewidth]{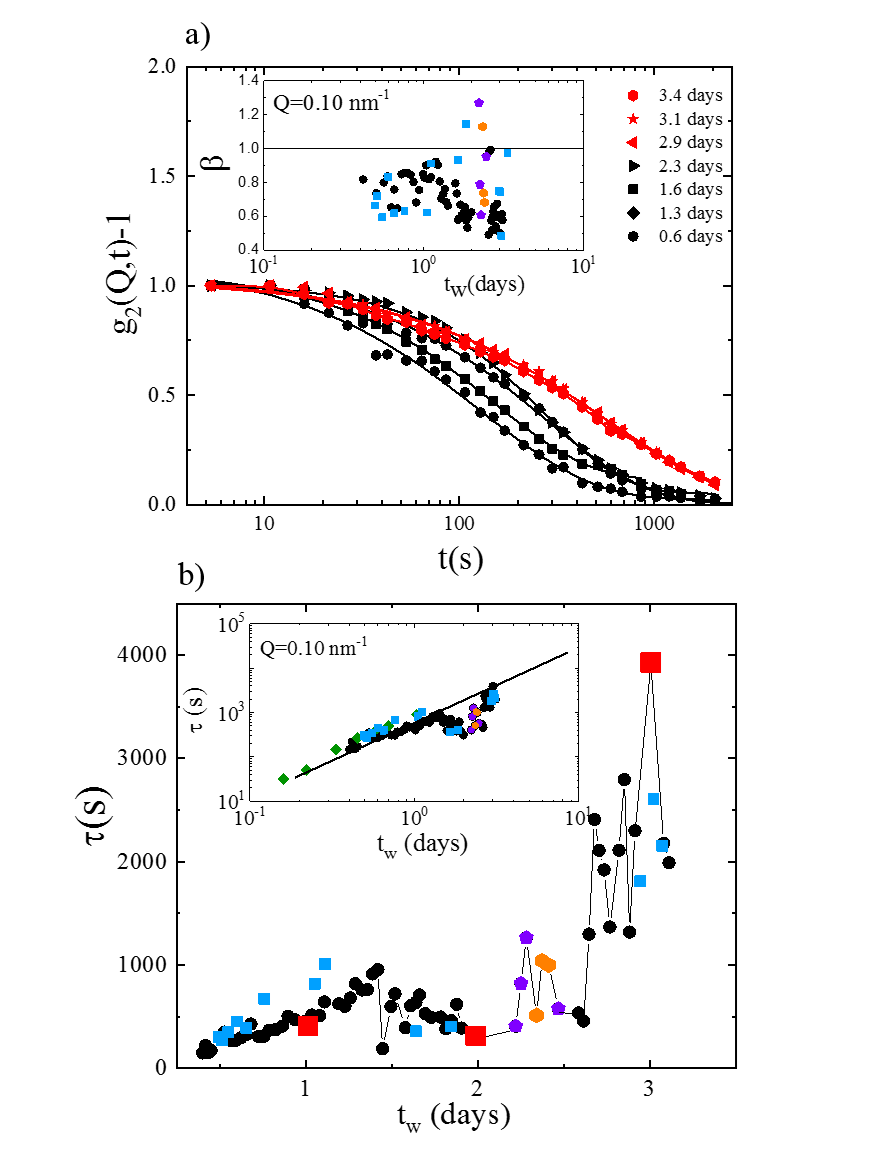}
\caption{
\textgm{(a) Some examples of XPCS normalized intensity autocorrelation
functions $g_2(Q,t)-1$ at wavevector Q = 0.10 nm$^{-1}$ and at several waiting 
times just across the glass-glass transition, as indicated.
Solid lines are \textit{stretched exponential} fits, as in 
Eq.~\ref{eq:stretched}.  Inset: \textit{stretching exponent}
$\beta$ as a function of waiting time for four different 
datsets (see Sec.~\ref{sec:XPCSmethods} for details).}
(b) Relaxation time $\tau$ as a function of waiting time for the same datasets as in panel a; red squares indicate the value of $\tau$ at one, two and three days.
Inset: data in the main panel are plotted in log-log scale, together with the relaxation time, reported in~\cite{BandyopadhyayPRL2004}, at a slightly larger wave-vector Q= 0.14 nm$^{-1}$ and in a range of relatively small $t_w$ (green squares). The solid line is a power-law guide-to-the-eyes, $\tau\propto t_w^{1.8}$.
}
\label{fig7}
\end{figure}

\textgm{\subsection{XPCS results}}
\textgm{Some examples of normalized intensity autocorrelation functions $g_2(Q,t)-1$ as a function of time and at fixed wave-vector\footnote{the value of $Q$ is somewhat smaller than the main peak of the structure factor in the WG~\cite{RuzickaPRL2010}, $Q^*=0.15 nm^{-1}$, and has been selected so as to have a good contrast for the scattered intensity.}, $Q = 0.10 nm^{-1}$, are shown in Fig.~\ref{fig7}a for Laponite samples at weight concentration C$_w$=3.0\% and at different waiting times. The time-dependence of the correlation functions is generally well described by the fitting expression:
\begin{equation}
\label{eq:stretched}
g_2(Q,t)-1 = A\exp[-(t/\tau)^\beta]^2 ~, 
\end{equation}
where $\tau$ is a characteristic relaxation time and $\beta$ is the
Kohlrausch exponent, also known as \textit{stretching exponent}.
}

The  relaxation time $\tau$ obtained through these fits 
is shown in Fig.~\ref{fig7}b as a function of the ageing time  in a range $0.5-3.2$ days, just across the suggested glass-glass transition. The figure features four different sets of data at  C$_w$=3.0\%, including a dataset from~\cite{AngeliniSM2013} (black circles), as well three new datasets, obtained for suspensions in water (violet \textgmm{pentagons} and orange hexagons) and in deuterated water D$_2$O (light blue \textgmm{squares}), respectively (see Sec.~\ref{sec:XPCSmethods} for details).
We have also marked with red squares the relaxation times at $t_w$ = 1, 2 and 3 days,  to highlight that the corresponding values mirrors the behavior observed for the rheological quantities, with a marked jump (about one order of magnitude) between two and three days. This result points to the existence of a correlation between relaxation time and the mechanical moduli.
Figure~\ref{fig7}b also shows the presence of a shallow minimum around $t_w$ = 2 days. \textgm{This feature seems to be quite robust and reproducible, as demonstrated by the coherent behaviour of the different datasets.} The inset in bi-logarithmic scale better illustrates that this minimum in fact takes the form of a downward deviation with respect to the power-law $\tau \propto  t_w^{1.8}$, clearly visible in a range of smaller waiting times. 
Notice that the inset also includes a dataset at a slightly larger wave-vector Q=0.14 nm$^{-1}$ from~\cite{BandyopadhyayPRL2004}, to show that the power-law scaling preceding the minimum is also quite robust. 
An analogous observation of a minimum in the waiting time dependence of the relaxation time had been previously reported for polystyrene colloids and attributed to the "first appearance" of large inhomogeneities in the sample~\cite{CipellettiPRL2000}.
We envisage that, in our samples, an onset of large  inhomogeneities is related to the occurrence of a glass-glass transition. \textgx{Moreover,} the high resolution (in terms of $t_w$) of our XPCS data reveals how, in the time-range around 2-3 days, the above discussed features of the relaxation time occurs on the top of quite large fluctuations, as evident in the main panel of Fig.~\ref{fig7}b.
The observation of such fluctuations 
provides
evidence  of an "anomalous" microscopic dynamics in Laponite\textsuperscript{\textregistered}, right in correspondence of the supposed glass-glass transition.

\textgm{The inset of Fig.~\ref{fig7}a shows the stretching exponent obtained from the same fits  (Eq.~\ref{eq:stretched}) as in Fig.~\ref{fig7}a, and points to a stretched exponential decay ($\beta<1$) of the 
correlation functions at almost all waiting
times.
In our spontaneously aged samples,  $\beta$ does not switch from  a stretched ($\beta< 1$) to a compressed ($\beta>1$) exponential behavior across the glass-glass transition, as reported for rejuvenated samples~\cite{AngeliniNC2014}. However, a “localized fluctuation” with $\beta>1$ is observed at waiting times corresponding to the dip of the relaxation time (inset of Fig 7b). It worth remarking that “stretched-to-compressed” transitions are commonly ascribed to the release of internal stresses in the sample, as observed in several colloidal systems~\cite{Bouchaud2008,ballesta2008unexpected,PastoreSciRep2017,NigroMacromol2020,pastore2019concentrated,PhilippePRE2018}. In the case of Laponite\textsuperscript{\textregistered}, such transition might be driven by the rejuvenation process through a syringe, as discussed in~\cite{AngeliniSM2013}. Here, in the lack of any rejuvenation, we speculate that the restructuring of correlated regions
from the first to the second glass is accompanied by bursts of internal stress releases, which tend to
cease after the transition, with an ensuing recovery of the stretched exponential behaviour.}

\textgm{Overall, our analysis indicates that signatures of the glass-glass transition can be spotted out through XPCS measurements even in the absence of any rejuvenation protocol, especially focusing on the relaxation time, and therefore nicely complements the conclusions of~\cite{AngeliniNC2014}, which instead focused on the stretching exponent of rejuvenated samples.}

\section{Discussion and Conclusions}
In this paper, we have studied  Laponite\textsuperscript{\textregistered} suspensions at C$_w$ = 3.0 \% in salt-free water, at different waiting times, and in the absence of any sample rejuvenation, through Dynamical Mechanical Analysis and XPCS.
Overall, the present results clearly support the existence of a spontaneous transition between two glasses with distinctive properties, up to the macroscopic rheological scale.
\textgm{The characteristic waiting time range (between 2 and 3 days) of the transition found in our work corresponds to that in~\cite{AngeliniNC2014}, strongly suggesting that the observed transition is the same. Another confirmation comes from our measurement of the moduli E$^{'}$ and E$^{''}$ being of the same order of magnitude of G$^{'}$ and G$^{''}$ reported in~\cite{AngeliniNC2014}.
The experimentally observed transition was previously interpreted as the change from a WG to a DHOC glass, drawing on hints from MC simulations~\cite{AngeliniNC2014}, and in the following we refer to this nomenclature (WG and DHOC) for convenience.}

In summary, all the investigated rheological quantities show a clear jump at an waiting time up to around three days.
Our findings include distinctive features of the two glasses, in terms of residual strain and elastic moduli upon deformation cycles and frequency sweeps, both in compression mode. Results point out that DHOC \textgx{glass} is  significantly less plastic and harder than the WG. \textgx{Moreover,} beyond the transition time, material properties seem to remain roughly constant up to the largest investigated age (one week), indicating that DHOC glass is quite stable.


Concerning dynamic XPCS measurements, our analysis firstly demonstrates that the glass-glass transition
can be detected by means of this technique, even without performing sample rejuvenation, as done in~\cite{AngeliniNC2014}.
Indeed, the relaxation time $\tau(t_w)$ determined through XPCS also shows a jump between 2 and 3 days of aging, hence correlating with the above jump in the mechanical quantities.
The high temporal resolution of these measurements (in terms of waiting time) allows us also to highlight for the first time the presence of severe fluctuations of $\tau$ emerging in the same time-range (2-3 days). 
We speculate that such fluctuations arise from an ongoing marked change of the structure of  Laponite\textsuperscript{\textregistered} suspensions from a glassy state to another glass,  which takes place in a correlated fashion.

\textgm{Generally, the structural relaxation of aging glassy materials is a correlated process, showing spotty bursts of activity, due to the cooperative motion of relatively large regions.}
Such kind of spatially correlated rearrangements typically generate large fluctuations of dynamical quantities \textgr{in a variety of soft glassy materials (see, e.g.s~\cite{DuriPRL2009,BouzidNatCom2017,song2022microscopic})}.
\textgm{Thus, we do expect that the restructuring underlying the present glass-glass transition occurs through cooperative rearrangements of large regions that spottily pass from a glassy state to another one, and are possibly facilitated by local stress release, as suggested by the temporary localized appearence of compressed exponential decay of $g_2-1$.
These changes may likely consist in  domains of particles switching from the local structure typical of WG to the T-shaped configurations of a DHOC, as suggested by the MC simulations of~\cite{AngeliniNC2014}.
The typical local relaxation time of such rearranging regions is expected to be smaller than the characteristic relaxation times of the two “stable” glassy states, and to affect the global relaxation time. Indeed, in this picture, the global relaxation time at a given $t_w$ would result from an average over the (different) local relaxation times corresponding to regions that are i) still in the first (WG) glass, ii) already in the second (DHOC) glass, and iii) actually undergoing a rearrangement from the first to the second glass.
Accordingly, the decreasing behaviour and the ensuing dip of the global relaxation time in proximity of the transition may be related to the first regions that start to undergo the transition, whereas the following fluctuations may be related to the later intermittent activation of the restructuring of other regions.}


In our opinion, the relevance of our work is twofold. On the one hand, our results are relevant for advancing the fundamental understanding  of the glass-glass transition, a distinctive phenomenon in soft glassy materials. The spontaneous glass-glass transition here found in Laponite\textsuperscript{\textregistered} suspensions is an example of a phenomenon that might occurr in other aging soft material, and therefore our results  open the way to the search for other systems undergoing this transition. Interesting examples may include colloidal rods~\cite{solomon2010microstructural}, as well as other clays and/or heterogeneously charged colloidal systems with competing attractive–repulsive electrostatic interactions. 
On the other hand, in general terms, the knowledge and control of aging processes occurring through different glassy states is crucial in technological applications, when assessing the long-term stability of soft materials.
In fact, nanosilicates, and in particular Laponite\textsuperscript{\textregistered}, are increasingly exploited as alternative systems to conventional 2D nanoparticles with enhanced physical, chemical and biological functionalities~\cite{XavierACSNano2015}. More specifically, the long-term stability of the system, discussed in the present paper, is of critical and vital importance for their highly promising applications in biomedical fields, such as bone tissue engineering~\cite{XavierACSNano2015,LokhandeActaBiomat2018,ShiAdvHMat2018} and 3D bioprinting fillers~\cite{AhlfeldBiofab2017, HeidActaBiomat2020}.

\textgr{As a future} perspective, it will be interesting to explore whether the  coupling between external load and hardening, which we here found to be quite evident in young samples of Laponite\textsuperscript{\textregistered}, may have more general implications. As an example, it is tempting to speculate that the glass-glass transition can be accelerated by persistent stress cycles.

\section*{Supplementary material}
See supplementary materials for additional figures, concerning sample preparations and reproducibility tests.

\begin{acknowledgments}
We acknowledge the European Synchrotron Radiation Facility (ESRF)
for provision of synchrotron radiation facilities, and we would like to
thank A. Fluerasu, A. Madsen and B. Ruta for assistance and support in using beamline ID10."
We  aknowledge also G. Franzino for assistance and support with DMA measurements.
\end{acknowledgments}

\section*{Author declarations}
\subsection*{Conflict of interest}
The authors have no conflicts to disclose.

\section*{Data Availability Statement}
The data that support the findings of this study are available from the corresponding authors upon reasonable request.

\bibliography{aipsamp}

\end{document}